\documentclass[11pt]{article}

\usepackage{indentfirst}
\usepackage{amssymb}
\usepackage[utf8]{inputenc}
\usepackage{mathptm}
\usepackage{ae}
\usepackage{bm}
\usepackage{amsthm,amsfonts,latexsym,fancyhdr,amsmath,tabularx}
\usepackage{graphicx}
\usepackage{subfigure, multicol}
\usepackage{epsfig}
\usepackage{verbatim}
\usepackage{pgf,tikz,pgfplots,gnuplottex}
\usepackage{multirow}
\usepackage{url,fancybox,color,makeidx}
\pgfplotsset{compat=1.17}
\usetikzlibrary{arrows}

\newtheorem{theorem}{Theorem}[section]

\newtheorem{definition}{Definition}[section]

\begin{document}

\title{Algebraic and Geometric Characterizations Related to the Quantization Problem of the $C_{2,8}$ Channel}

\author{Anderson José de Oliveira
\thanks{Department of Mathematics, Institute of Exact Sciences, Federal University of Alfenas (UNIFAL-MG), 37130-000, Alfenas, MG, Brazil,
E-mail:~{\tt \small anderson.oliveira@unifal-mg.edu.br}.
}
Giuliano Gadioli La Guardia
\thanks{Department of Mathematics and Statistics,
State University of Ponta Grossa (UEPG), 84030-900, Ponta Grossa, PR, Brazil,
E-mail:~{\tt \small gguardia@uepg.br.}
}
Reginaldo Palazzo Jr. \\
\thanks{Department of Communications, FEEC-UNICAMP, 13083-852, Campinas, SP, Brazil,
E-mail:~{\tt \small palazzo@dt.fee.unicamp.br.}
}
Clarice Dias de Albuquerque 
\thanks{Science and Technology Center, Federal University of Cariri (UFCA), 63048-080, Juazeiro do Norte, CE, Brazil,
	E-mail:~{\tt \small clarice.albuquerque@ufca.edu.br}
}
Cátia Regina de Oliveira Quilles Queiroz \\
\thanks{Department of Mathematics, Institute of Exact Sciences, Federal University of Alfenas (UNIFAL-MG), 37130-000, Alfenas, MG, Brazil,
E-mail:~{\tt \small catia.quilles@unifal-mg.edu.br.}
}
Leandro Bezerra de Lima
\thanks{Department of Mathematics, Federal University of Mato Grosso do Sul (UFMS), 79200-000, Campo Grande, MS, Brazil,
	E-mail:~{\tt \small leandro.lima@ufms.br }
}
Vandenberg Lopes Vieira 
\thanks{Department of Mathematics, State University of Paraíba (UEPB), 58429-500, Campina Grande, PB, Brazil,
E-mail:~{\tt \small vandenberglv@uepb.edu.br.}
}
}

\maketitle

\begin{abstract}
\noindent In this paper, we consider the steps to be followed in the analysis and interpretation of the quantization problem related to the $C_{2,8}$ channel, where the Fuchsian differential equations, the generators of the Fuchsian groups, and the tessellations associated with the cases $g=2$ and $g=3$, related to the hyperbolic case, are determined. In order to obtain these results, it is necessary to determine the genus $g$ of each surface on which this channel may be embedded. After that, the procedure is to determine the algebraic structure (Fuchsian group generators) associated with the fundamental region of each surface. To achieve this goal, an associated linear second-order Fuchsian differential equation whose linearly independent solutions provide the generators of this Fuchsian group is devised. In addition, the tessellations associated with each analyzed case are identified. These structures are identified in four situations, divided into two cases $(g=2$ and $g=3)$, obtaining, therefore, both algebraic and geometric characterizations associated with quantizing the $C_{2,8}$ channel.
\end{abstract}

\section{Introduction}
\begin{figure}[!h]
	\centering
	\includegraphics[scale=0.7]{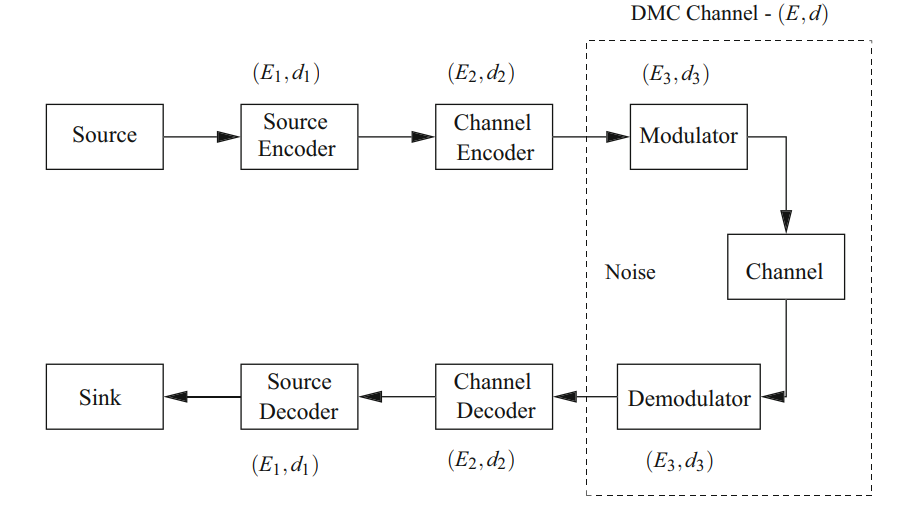}
	\caption{Communication system model, from \cite{rodrigo,canal}.}
	\label{channel}
\end{figure}

The goals of designing a new digital communication system are to construct more reliable and less complex systems than previously known ones. To achieve these goals, the concept of a metric space $(E_i, d_i)$ associated with each block in the traditional communication system model is employed, as shown in Figure~\ref{channel}.

The use of the bit error probability measures the reliability of a digital communication system. Therefore, the system is more reliable if its bit error probability is the least bit error probability among the previously known systems, whereas to be less complex, it is required that the signal set be geometrically uniform, \cite{forney}. As shown in \cite{rodrigo}, these goals may be achieved if the new approach makes use of an essential topological invariant known as the \textit{genus} $g$ of an oriented compact surface where the set of points (signal set) lies. Since the channel is discrete and memoryless, this channel may be seen as a complete bipartite graph. Thus, one way to obtain the genus of the surface is to find the \textit{2-cell} embedding of the complete bipartite graph on the corresponding oriented compact surface.

Most of the communication channels of practical interest, which are associated with complete bipartite graphs, may be embedded on surfaces with genus greater than or equal to 2. Therefore, the geometry to be considered is non-Euclidean geometry. On the other hand, it is well-known that the geometric uniformity of a signal set $S$ is a property associated with its group of isometry. If the genus is greater than or equal to two, the extension of this concept to the hyperbolic plane implies knowing the hyperbolic isometries and their geometric properties.

On the other hand, Massey in \cite{massey} has shown that under the error probability criterion, the performance of a binary digital communication system using soft-decision in the demodulator, for instance, an $8$-level quantizer, leading to a binary input, $8$-ary output symmetric channel, denoted by $C_{2,8}$, achieves a gain of up to 2 dB when compared to the performance of a binary digital communication system using hard-decision (a $2$-level quantizer, leading to a binary symmetric channel (BSC), denoted by $C_{2,2}$).

How does one mathematically justify this performance improvement? By performing a 3-bit quantization (8-level quantizer), Cavalcante \textit{et al.}, \cite{rodrigo}, have shown that the associated graph corresponding to this channel may be embedded on surfaces with genus $0 \le g \le 3$ with $g=3$ being the case achieving the 2 dB gain as mentioned earlier. Based on this fact, we may argue that the surface genus may be taken as a new reliability measure of a digital communication system from the topological point of view.

In addition, the signal design problem under the geometric uniformity point of view asks for finding the isometry group associated with a signal constellation on the surface provided by the topological point of view. One way to determine the uniformity of the decision region associated with each signal in the signal constellation is to use a particular class of linear ordinary differential equations, the so-called \textit{Fuchsian differential equations}. The determination of the uniformity of the decision region of a signal in a signal constellation is equivalent to finding the uniformization region of a given algebraic curve.

Fuchsian differential equations represent an important class of linear ordinary differential equations whose main characteristic is that every singular point in the extended complex plane is regular. These differential equations are widely used in Mathematical Physics problems. The most studied cases involve equations with three regular singular points, such as the hypergeometric, Legendre, and Tchebychev equations. In comparison, the Heun equation contains four regular singular points.

In \cite{canal}, a relationship between the singularities of Fuchsian differential equations and hyperbolic geometry is considered. The three regular singular points of the hypergeometric equation, $\{0,1,\infty\}$, determine the fundamental region of $\Gamma_0$, the associated Fuchsian group, as a hyperbolic triangle whose genus is $g = 0$. In addition, an analysis of the connection with the channel quantization problem for transmitting information in digital communication systems is realized.

On the other hand, when considering a Fuchsian differential equation with five or more singular points, \cite{canal,alves} have shown that the genus of the uniformization region of the algebraic curve and the topological genus of the fundamental region are not equal due to the existence of an elliptical or parabolic transformation between the edge-pairings in the fundamental polygon. Thus, a constraint exists associated with the number of singularities to achieve equality between the corresponding genus. Therefore, to the best of our knowledge, an alternative for the uniformization of algebraic curves of degrees greater than or equal to 5 is to employ Whittaker's procedure to achieve the genus equality.

In \cite{Whittaker, Whittaker2, Mursi}, relevant connections are made among Fuchsian differential equations, Riemann surfaces, and Fuchsian groups
associated with them to analyze the uniformization of the algebraic curves of the form $y^2=z^{2g+1}\pm 1$.

The procedure proposed by Whittaker, \cite{Whittaker}, aims to find the uniformization region of the algebraic curve $y^2 = z^5 + 1$. In this case, the coefficient associated with the highest order derivative of the corresponding Fuchsian differential equation is $z^5 + 1$. The linearly independent solutions of this differential equation lead to the five elliptic transformations associated with each side of a regular pentagon and, therefore, to the generators of the Fuchsian group $\Gamma_0$. Fixing one of these Möbius transformations and multiplying it by the remaining ones leads to an eight sides regular polygon, the uniformization region of the given algebraic curve. The generators of this fundamental region are hyperbolic Möbius transformations associated with the Fuchsian group $\Gamma_{8}$. As a consequence, the associated Riemann surface is the 2-torus.

Mursi in \cite{Mursi}, considered a similar procedure as the one proposed by Whittaker, \cite{Whittaker}, to find the uniformization region of the algebraic curve $y^2 = z^7 + 1$. In this case, the coefficient associated with the highest order derivative of the corresponding Fuchsian differential equation is $z^7 + 1$. The linearly independent solutions of this differential equation lead to the seven elliptic transformations associated with each side of a regular heptagon and, therefore, to the generators of the Fuchsian group $\Gamma_0$. Fixing one of these Möbius transformations and multiplying it by the remaining ones leads to a twelve sides regular polygon, the uniformization region of the given algebraic curve. The generators of this fundamental region are hyperbolic Möbius transformations associated with the Fuchsian group $\Gamma_{12}$. As a consequence, the associated Riemann surface is the 3-torus.

As mentioned previously, the $C_{2,8}$ channel may be embedded on oriented compact surfaces with genus ranging from zero to three, that is, $0\le g \le 3$, see \cite{rodrigo} for more detailed information. For genus $g=0,1$, the geometry is Euclidean. The geometric and algebraic structures of the Euclidean case are well-known and presented in \cite{canal}. The geometry is hyperbolic when the genus is $g=2$ and $g=3$.

In this paper, we consider the four algebraic curves (hyperelliptic curves) given by $y^2 = z^{2g+1} - 1$ and $y^2 = z^{2g+2} - 1$, with $g=2$ and $g=3$, as follows:

\begin{itemize}
	\item for genus $g=2$: the hyperelliptic curve is given by $y^{2}=(z^{5}-1)$, leading to the $n=5$ symmetric points $s_1, s_2, \ldots, s_{5}$;
	\item for genus $g=2$: the hyperelliptic curve is given by $y^{2}=(z^{6}-1)$, leading to the $n=6$ symmetric points $s_1, s_2, \ldots, s_{6}$;
	\item for genus $g=3$: the hyperelliptic curve is given by $y^{2}=(z^{7}-1)$, leading to the $n=7$ symmetric points $s_1, s_2, \ldots, s_{7}$;
	\item for genus $g=3$: the hyperelliptic curve is given by $y^{2}=(z^{8}-1)$, leading to the $n=8$ symmetric points $s_1, s_2, \ldots, s_{8}$.
\end{itemize}

It is worth mentioning that the Fuchsian groups corresponding to the uniformization regions of the algebraic curves $y^{2}=(z^{5}-1)$ and $y^{2}=(z^{6}-1)$, both with genus 2 are not isomorphic, as well as the Fuchsian groups corresponding to the uniformization regions of the algebraic curves $y^{2}=(z^{7}-1)$ and $y^{2}=(z^{8}-1)$. Although the hyperbolic area of the uniformization region is the same for each $g$, that is, $4\pi$ when $g=2$, and $8\pi$ when $g=3$, the number of neighbors are different, for the corresponding tessellations are $\{8,8\}$ and $\{10,5\}$. Therefore, from a practical point of view, there is a preference for employing the tessellation $\{10,5\}$ due to its reduced number of neighbors, leading to a lower error probability. This is one of many cases that may be considered. For instance, it could be applied in different situations, such as searching for better-performing codes than those already known by using generalized concatenated codes, and so on.

This paper aims to present the algebra (Fuchsian groups generators) and geometry (polygon side-pairings and associated tessellations) related to the cases $g=2$ and $g=3$ in order to analyze the channel $C_{2, 8}$ quantization problem, which can be embedded in surfaces of genus $0 \leq g \leq 3$, through a second order Fuchsian differential equation.

This paper is organized as follows. In Section 2, the main concepts related to graphs and surfaces, hyperbolic geometry, Fuchsian differential equations, and the uniformization of the algebraic curves, with the procedure proposed by Whittaker and Mursi to identify the generators of the Fuchsian group are presented. In Section 3, we present the results of this paper by considering the four cases (two for genus two and two for genus 3). Finally, in Section 4, we discuss the final remarks of the paper.

\section{Preliminaries} \label{Sec2}
This section recalls some general concepts necessary for developing this paper.

\subsection{Graphs and surfaces}\label{subsec21}

\begin{definition}
	\cite{White1} A graph $G$ consists of a finite and non-empty set $V(G)$ of elements called
	vertices and a set $A(G)$ of elements called edges, which are unordered pairs
	of vertices.
\end{definition}

\begin{definition}
	\cite{White} A graph $G$ is called an \textit{embedding} in a surface $\Omega $ when no two of its edges meet except at a vertex.
\end{definition}

The complement of $G$ in $\Omega $ is called \textit{region}. A region that is homeomorphic to an open disk is called \textit{2-cell}; if the entire region is a \textit{2-cell}, the embedding is said to be a \textit{\textit{2-cell} embedding}. It is known that if $G$ is connected, then the minimum embedding is a \textit{2-cell} embedding.

\begin{definition}
	\cite{White} A \textit{complete bipartite graph} with $m$ and $n$ vertices, denoted by $K_{m,n},$ is a graph consisting of two disjoint vertex sets with $m$ and $n$ vertices, respectively, where an edge connects each vertex of a set to every vertex of the other set.
\end{definition}

When considering \textit{\textit{2-cell}} embedding of complete bipartite graphs $K_{m,n}$, the minimum and the maximum genus of the corresponding surfaces may be determined. These values are given in the following:

\begin{enumerate}
	\item[$\left( i\right) $]  the minimum genus of an oriented compact surface is, \cite{Ringel}:
	\begin{equation}
		g_{m}\left( K_{m,n}\right) =\{\left( m-2\right) \left( n-2\right)/4\},\;\; \mbox{for }m,n\geq 2,  \label{gen min kmn}
	\end{equation}
	where $\{a\}$ denotes the least integer greater than or equal to the real number $a$;
	
	\item[$\left( ii\right) $]  the maximum genus of an oriented compact surface is, \cite {Ringeisen}:
	\begin{equation}
		g_{M}\left( K_{m,n}\right) =\left[ \left( m-1\right) \left(n-1\right) /2\right] ,\;\;\mbox{for }m,n\geq 1,  \label{gen mas
			kmn}
	\end{equation}
	where $\left[ a\right] $ denotes the greatest integer less than or equal to the real number $a$.
\end{enumerate}

\subsection{Review of hyperbolic geometry}

The genus $g$ of a compact orientable surface $\mathbb{M}$ is determined by the number of handles connected to a sphere or the number of ``holes'' in $\mathbb{M}$. As the Euler characteristic of a surface, the genus is a topological invariant; that is, it is the same for homeomorphic surfaces. We can identify surfaces with genus $g=1$ with the torus, and the geometry associated with these surfaces is Euclidean geometry. For surfaces of genus $g \geq 2$, $g$-tori, the geometry to be considered is the {\em hyperbolic geometry}. The difference between hyperbolic and Euclidean geometries is that the former fails to satisfy Euclid's fifth postulate, the axiom of parallelism. This section reviews some basic concepts of hyperbolic geometry necessary for developing this paper. We suggest the references \cite{Katok1992,Firby1991,Beardon1983,Stillwell2000} for a more in-depth review of this subject.

We consider two models of hyperbolic geometry: the {\em upper half-plane}, $\mathbb{H}^2 = \{z \in \mathbb{C} \ | \ Im(z) > 0\}$, and the {\em Poincar\'e disk} $\mathbb{D}^2 = \{z \in \mathbb{C} \ | \ |z| < 1\}$. In addition to these models, there are the Klein and hyperboloid models, which we do not discuss here.

An advantage of the Poincar\'e disk model over the upper half-plane model is that the unit disk $\mathbb{D}$ is a bounded subset of the Euclidean plane. On the other hand, the upper half-plane model is better for performing computation using cartesian coordinates.

The {\em hyperbolic plane} is the space $\mathbb{H}^2$ equipped with the hyperbolic metric:
\begin{equation} \label{eq1}
	ds = \frac{\sqrt{dx^2 + dy^2}}{y}.
\end{equation}

Another fundamental concept is {\em geodesic}, a path with the shortest hyperbolic length connecting two distinct points. The geodesics in $\mathbb{H}^2$ are half circles and half lines orthogonal to the real axis $\mathbb{R} = \{z \in \mathbb{C} \ | \ Im(z) = 0\}$, while the geodesics in $\mathbb{D}^2$ are segments of Euclidean circles orthogonal to $\partial \mathbb{D}^2$; in particular their diameters. A single geodesic can connect any two points $z,w \in \mathbb{H}^2$.

The {\em hyperbolic angle} between two geodesics in $\mathbb{H}^2$ intersecting at a point $z$ is the (Euclidean) angle between the tangent vectors to the geodesics. Recall that the sum of the interior angles of a hyperbolic triangle is less than $\pi$.

The Gauss-Bonnet theorem shows that the hyperbolic area of a hyperbolic triangle depends only on its angles.

\begin{theorem} (Gauss-Bonnet)
	Let $\Delta$ be a hyperbolic triangle with interior angles $\alpha, \beta, \theta$. Then the area of $\Delta$ is given by
	$$\mu (\Delta) = \pi - \alpha - \beta - \theta.$$
\end{theorem}

A {\em hyperbolic polygon} (or a $p$-gon) P with $p$ sides is a closed convex set consisting of $p$ segments of hyperbolic geodesics. The intersection of two geodesics is called {\em vertex} of the polygon. Such a polygon whose sides have the same length and the interior angles are equal is called a regular polygon.

We are interested in the side pairing of a hyperbolic polygon, performed by elements of a Fuchsian group, which we define in the sequence.

The {\em unimodular group} $SL(2, \mathbb{R})$ is the multiplicative group of real matrices $2\times 2$:
$$A = \left(
\begin{array}{cc}
	a & b \\
	c & d \\
\end{array}
\right),$$  \label{matriz}
where $a, b, c, d \in \mathbb{R}$ and $\det(A) = 1$.

\begin{definition}
	A {\em Möbius transformation} is a map $T: \mathbb{C} \rightarrow \mathbb{C}$ given by:
	$$T(z) = \frac{az + b}{cz + d},$$
	where $a, b, c, d \in \mathbb{R}$ and $ad - bc = 1$.
\end{definition}
We can represent a Möbius transformation using matrices of the type $\pm A$, where $A \in SL(2,\mathbb{R})$. We thus define the {\em linear special projective group}, denoted by $PSL(2, \mathbb{R})$, as the multiplicative group of Möbius transformations; equivalently, $PSL(2, \mathbb{R})\cong SL(2,\mathbb{R})/\langle \pm I_2 \rangle$.

Möbius transformations are divided into three distinct classes: {\em elliptic}, {\em parabolic}, and {\em hyperbolic}. The classification of these transformations depends on the {\em trace function of $T$} defined by $Tr(T) = [tr(A)]^2 = [tr(-A)]^2$, where $T$ is a Möbius transformation corresponding to the pair of matrices $\pm A \in SL(2, \mathbb{R})$ and $tr(A)$ is the usual trace function of matrices. Hence, for $T \in PSL(2, \mathbb{R})\setminus {I_2}$, one has:
\begin{itemize}
	\item[(i)] $T$ is elliptical if and only if $Tr(T) < 4$;
	\item[(ii)]$T$ is parabolic if and only if $Tr(T) = 4$;
	\item[(iii)] $T$ is hyperbolic if and only if $Tr(T) > 4$.
\end{itemize}

The Möbius transformations are homeomorphisms and preserve the hyperbolic distance in $\mathbb{H}^2$. Thus, the group $PSL(2,\mathbb{R})$ is a subgroup of the group of all isometries of $\mathbb{H}^2$. Consequently, any transformation into $PSL(2, \mathbb{R})$ takes geodesic into geodesic.

\begin{definition}
	A subgroup $\Gamma \subset PSL(2, \mathbb{R})$ is discrete if the topology induced on $\Gamma$ is a discrete topology, that is if $\Gamma$ is a discrete set in the topological space $PSL(2, \mathbb{R})$.
\end{definition}

\begin{definition}
	A Fuchsian group is a discrete subgroup of $PSL(2, \mathbb{R})$.
\end{definition}

A {\em regular tessellation} of the hyperbolic plane is a partition consisting of congruent polygons, subject to the constraint of intercepting only at edges and vertices, to have the same number of polygons sharing the same vertex, independent of the vertex.

From the Gauss-Bonnet Theorem, a regular hyperbolic $\{p,q\}$ tessellation must satisfy the inequality $(p - 2)(q - 2) > 4$. Consequently, there exist infinite regular tessellations on the hyperbolic plane.

Under certain conditions, a compact topological surface $M$ can be obtained by pairing the sides of a polygon $P$. As long as the lengths of the sides are equal, a side pairing transformation is an $\gamma \neq Id$ isometry of an isometry group that preserves orientation $\Gamma$, taking one side $s$ of $P$ to another side $\gamma (s) = s'$ of $P$, and also, $\gamma^{-1} \in \Gamma \setminus \{Id\}$ takes $\gamma (s) = s'$ to $s$. If $s$ is identified with $s'$, and $s'$ is identified with $s''$, then $s$ is identified with $s''$. Such an identification chain can also occur with vertices: we call a maximal set $\{v_1, v_2, \ldots v_k\}$ of identified vertices a {\em vertex cycle}. When the angles of each vertex cycle add up to $2\pi$, the identification space of the paired sides of $P$ results in a hyperbolic surface.

Note that the number of sides of $P$ is even since the sides are identified in pairs. However, it is possible to have an odd number of sides, in which case a side must be identified with itself by adding a vertex in the middle of that side.

\subsection{Ordinary differential equations in the complex domain}

In many cases, a more detailed study of the properties of an ordinary differential equation (ODE) and its solutions requires using the complex plane.

The concepts and main results presented in the following can be found in \cite{Gerhard, soto, Metodos}.

Consider the ODE $\frac{dx}{dy}=-y^2$, whose solutions are of the form $y(x)=(x-c)^{-1}$, where $c$ is an arbitrary real constant. We must consider two regions: $x>c$ and $x<c$. There is no way to go from one region to another, as the only path along the line passes through the point $x=c$, which must be excluded. It can also be observed that the behavior of the solutions is different for $x\rightarrow c^{\pm}$, i.e., $\lim_{x\rightarrow c^{\pm}}y(x)=\pm \infty$. On the other hand, if we consider complex variables, we have the ODE $\frac{dw}{dz}=-w^2$, which has solutions $w(z)=(z-c_1)^{-1}$, where $c_1$ is, in general, an arbitrary complex constant, but let us consider $c_1=c$. In this way, we do not need to consider two separate regions but exclude the point $z=c$ as a singularity from the equation. Therefore, it is simpler to analyze the singularities of the ODE solutions when working in the complex plane.

\subsubsection{Fuchsian differential equations}

The study of Fuchsian differential equations aroused greater interest in mathematical researchers in the second half of the 19th century and the beginning of the 20th century, providing many developments in the theory of functions of complex variables.

Let us consider the complex plane with the addition of the point at infinity, i.e., the extended complex plane $\overline{\mathbb{C}}=\mathbb{C}$ $\cup$ $\infty$.

An equation of the type:
\begin{equation}
	y^{(n)}(z)+ p_1(z)y^{(n-1)}(z)+...+p_{n-1}(z)y^{'}(z)+p_n(z)y(z)=0,
\end{equation}
is a Fuchsian equation or an equation of the \textit{Fuchsian} type if every singular point in the extended complex plane is regular.

In the case of the second-order equation:
\begin{equation}
	y^{''}(z)+p_1(z)y^{'}(z)+p_2(z)y(z)=0,
\end{equation}
a singular point is said to be regular if the singularity at $p_1(z)$ is a simple pole and at $p_2(z)$ is a pole of order at most 2.

\begin{definition}
	A second-order Fuchsian differential equation with $n$ singular points is of the form:
	
	\begin{equation}y^{''}(z)+p_1(z)y^{'}(z)+p_2(z)y(z)=0,\end{equation}
	\noindent where:
	\begin{eqnarray*}
		p_1(z)&=&\frac{A_1}{z-\xi_1}+\cdots+\frac{A_n}{z-\xi_n}+K_1,\\
		p_2(z)&=&\frac{B_1}{(z-\xi_1)^2}+\frac{C_1}{z-\xi_1}+\cdots+\frac{B_n}{(z-\xi_n)^2}+\frac{C_n}{z-\xi_n}+K_2,
	\end{eqnarray*}
	
	\noindent and $A_i, B_i, C_i, i=\{1, \cdots, n\}$, $K_1$ and $K_2$ are complex constants, satisfying the following constraints:
	\begin{eqnarray*}
		A_1+\cdots+A_n&=&2,\\
		C_1+\cdots+C_n&=&0,\\
		(B_1+\cdots+B_n)+(\xi_1C_1+\cdots+\xi_nC_n&=&0,\\
		(2\xi_1B_1+\cdots+2\xi_nB_n)+(\xi_1^{2}C_1+\cdots+\xi_n^{2}C_n)&=&0.
	\end{eqnarray*}
	
\end{definition}

Fuchsian differential equations represent a type of ordinary differential equation with particular features and important applications in several areas, such as Mathematical Physics problems, in addition, to representing an important tool in the development of functions of complex variables.

The Legendre, Tchebychev, hypergeometric, and Heun equations are examples of Fuchsian differential equations. An important property related to these equations is that they can be solved around their singular points by the Frobenius method. Additionally, they have some transformation properties that facilitate their analysis, interpretation, and application. The main characteristics of these equations are presented below.

The Legendre equation is given by (\ref{eq7}):

\begin{equation}\label{eq7}
	y^{\prime \prime}(z)+\frac{2z}{1-z^2}y^{\prime}(z)+\frac{\lambda(\lambda+1)}{1-z^2}y(z)=0,
\end{equation}
where $\lambda \in \mathbb{C}$. The points $z_0=\pm1$ and $z_0=\infty$ are regular singular points of this equation.

The Tchebychev equation is given by (\ref{eq8}):

\begin{equation}\label{eq8}
	y^{\prime \prime}(z)+\frac{z}{1-z^2}y^{\prime}(z)+\frac{\lambda^2}{1-z^2}y(z)=0,
\end{equation}
where $\lambda \in \mathbb{R}$. The points $z_0=\pm1$ and $z_0=\infty$ are regular singular points of this equation.

The Heun equation is given by (\ref{eq9}):
\begin{equation}\label{eq9}
	y^{\prime \prime}(z)+ \left[\frac{\gamma}{z}+\frac{\delta}{z-1}+\frac{\epsilon}{z-a}\right]y^{\prime}(z)+\frac{\alpha \beta z-q}{z(z-1)(z-a)}y(z)=0,
\end{equation}
where $\alpha, \beta, \gamma, \delta, \epsilon$, $a$ and $q$ are complex constants. The points $z_0=0$, $z_0=1$, $z_0=a$, and $z_0=\infty$ are regular singular points of this equation.

In this paper, the Fuchsian differential equation that we will use is the hypergeometric equation, given by (\ref{eq10}) in the sequence, due to a series of characteristics and properties associated with the uniformization process of algebraic curves, presented in the next section.

\begin{itemize}
	\item {\bf{Hypergeometric equation}}
	\begin{equation}\label{eq10}
		y^{\prime \prime}(z)+\frac{[c-(1+a+b)z]}{z(1-z)}y^{\prime}(z)-\frac{ab}{z(1-z)}y(z)=0,
	\end{equation}
	where $a, b$ and $c$ are complex constants. The points $z_0=0$, $z_0=1$, and $z_0=\infty$ are regular singular points of the hypergeometric equation.
	
	Several differential equations of interest can be transformed into hypergeometric equations. These equations are (unless trivially multiplied by a constant) the only second-order homogeneous linear equation with only three regular singular points at $z_0=0$, $z_0=1$, and $z_0=\infty$.
\end{itemize}

It is possible to demonstrate that all Fuchsian differential equations with three singularities can be transformed into a hypergeometric equation, and with four singularities can be transformed into a Heun equation.

\subsection {Uniformization of algebraic curves }
In this section, it will be presented some essential concepts related to the uniformization of algebraic curves. We refer the reader to \cite{canal, Ford, Whittaker, Whittaker2, Dalzell, Mursi} for more detailed information on the concepts and main results shown below.

Consider next the circle $w^2+z^2=1$ represented in the parametric form as $z=\sin y$, and $w=\cos y$. Another parametric representation of the same curve is $z=\frac{2y}{1+y^2}$, and $w=\frac{1-y^2}{1+y^2}$. These two parameterizations are said to make the {\em uniformization} of the function $w^2+z^2=1$. Formally, one has:

\begin{definition} \cite{Ford}
	Let $w=f(z)$ be a multiple-valued function of $z$ and $w=w(z)$, $z=z(y)$ two non-constant single-valued functions of $y$, such that $w(y)=f(z(y))$. We say that $w$ and $z$  uniformize the $w=f(z)$ function. The variable $y$ is said to be a variable of uniformization.
\end{definition}

\begin{definition}\cite{Ford} A hyperelliptic curve is an algebraic curve given by an equation of the form $y^2=f(z),$ where $f(z)$ is a polynomial of degree $n>4$ with $n$ distinct roots. A hyperelliptic function is an element of the field of functions of a curve or, eventually, of the Jacobian variety in the curve.
\end{definition}

It is possible to identify the tessellation type of the hyperbolic plane using the following relations: if the degree of the algebraic curve is odd, that is, if $\partial f=2g+1$, where $g$ is the genus of the algebraic curve and also of the surface, the resulting tessellation is of the type $\{4g,4g\}$, whereas if the algebraic curve has even degree, that is, $\partial f=2g+2$, the resulting tessellation is of the type $\{4g+2, 2g+1\}$ \cite{canal}.

Whittaker and Mursi \cite{Whittaker}, \cite{Mursi} established a relationship between the coefficient of the higher order derivative of Fuchsian differential equations with Riemann surfaces to uniformize the planar algebraic curve. These works presented functions of the type $y^2=z^{2g+1}\pm 1$, where $g$ denotes the genus of the surface whose singularities can be identified as the roots of unity.

The generators of the Fuchsian group $\Gamma_0$ are derived as the quotient of two linearly independent solutions of the hypergeometric differential equation. These group generators determine the fundamental region of $\Gamma_0$ where each singularity is seen as a vertex of the fundamental polygon \cite{canal}. A corresponding elliptic transformation applied to each midpoint of the edges of this polygon generates the Fuchsian group $\Gamma_0$. By fixing one of these edges and multiplying it by the remaining ones, the generators of the Fuchsian group ($\Gamma_{4g}$ or $\Gamma_{4g+2}$) are determined. The resulting fundamental polygon uniformizes the given algebraic curve.

According to \cite{Whittaker, Whittaker2, Dalzell, Mursi}, the uniformization variable of the algebraic curve $y^2=(z-e_1)\cdot (z-e_2) \cdots (z-e_{2g+2})=f(z)$ is the quotient of two linearly independent solutions of the linear differential equation:

\begin{equation}\label{eq11}
	\frac{d^2y}{dz^2}+\frac{3}{16}\left\{\sum_{r=1}^{2g+2}\frac{1}{(z-e_r)^2}+\frac{(-2g+2)\cdot z^{2g}+2\cdot g \cdot p_1 \cdot z^{2g-1}+c_1\cdot z^{2g-2}+c_{2g-1}}{(z-e_1)\cdot (z-e_2) \cdots (z-e_{2g+2})}\right \}y=0,
\end{equation}
where $p_1= \sum e_r$ and $c_1, c_2, c_{2g-1}$ are constants (depending on $e_1, e_2, \cdots e_{2g+2}$), that are
determined by the condition that the group resulting from (\ref{eq11}) is Fuchsian.

Equation (\ref{eq11}) can be rewritten, according to \cite{Whittaker, Whittaker2, Dalzell, Mursi} as:

\begin{equation}
	\frac{d^2y}{dz^2}+  \frac{3}{16}\left[\left\{\frac{f^{\prime}(z)}{f(z)}\right\}^2-\frac{2g+2}{2g+1}\frac{f^{\prime \prime}(z)}{f(z)}\right]y=0.
\end{equation}

{\em Automorphic functions} are a natural generalization of circular and elliptical functions, \cite{Ford}. The importance of automorphic functions is because, through them, the uniformity problem can be solved; that is, if $y$ is an algebraic function of $z$, defined by an algebraic equation $f(y, z) = 0$, then $y$ and $z$ can be expressed by univariate functions of a third variable $t$ and these functions are automorphic. If the equation $f(y, z) = 0$ is of genus zero, the automorphic functions are rational algebraic functions or circular functions; when $f(y, z) = 0$ is of genus one, the
automorphic functions are elliptic functions, and when the genus of $f(y, z) = 0$ is greater than one, proper automorphic functions are needed. In the case of Whittaker's approach, the genus is at least two. Any algebraic equation of genus two can be represented by a birational transformation into hyperelliptic form, where $y^2$ is a polynomial of degree five or six in $z$. When we analyze algebraic equations of genus three, the polynomial is of degree seven or eight.

Whittaker \cite{Whittaker} showed for the case $y^2=(z-e_1)\cdot (z-e_2)\cdot (z-e_3)\cdot (z-e_4)\cdot (z-e_5)$, which represents the normal form for genus $2$, that the uniformization variable is the quotient of two linearly independent  solutions of the linear differential equation:

\begin{equation}\label{eq13}
	\frac{d^2y}{dz^2}+  \frac{3}{16}\left[\left\{\frac{f^{\prime}(z)}{f(z)}\right\}^2-\frac{6}{5}\frac{f^{\prime \prime}(z)}{f(z)}\right]y=0,
\end{equation}
where $c_1, c_2$ and $c_3$ are completely determined.

Through some transformations, which can be analyzed in detail in \cite{Whittaker}, (\ref{eq13}) can be transformed in a hypergeometric equation:

\begin{equation}\label{eq14}
	25x(x-1)\frac{d^2y}{dx^2}+20(2x-1)\frac{dy}{dx}+2y=0.
\end{equation}

Consider $y_1$ and $y_2$ as two independent solutions of (\ref{eq14}); hence, when $x$ describes any closed path in the $x$-plane with one or more singularities of the equation, the solutions $y_1$ and $y_2$ become, respectively, $ay_1+by_2$ and $cy_1+dy_2$, where $a, b, c, d$ are constants and the quotient $\frac{ay_1+by_2}{cy_1+dy_2}$ becomes $\frac{at+b}{ct+d}$ when $t=\frac{y_1}{y_2}$.

An algorithm is presented in \cite{Erika} to obtain the subgroup $G$ from the fundamental region of the orientable compact surface that uniformizes the planar hyperelliptic curve. In general terms, the steps to be followed are as follows:

\begin{itemize}
	\item Step 1 - Given a hyperelliptic curve, determine its roots;
	\item Step 2 - Determine the geodesics associated with subsequent roots as well as the midpoints of these geodesics;
	\item Step 3 - For each side of the hyperbolic polygon, determine the elliptic transformation: $S_r(t) = \frac{at+b}{ct+d}$, where $ad-bc=1$, the trace of $S_r<2$ and $r=1, \cdots, n$;
	\item Step 4 - Verify if $(S_rS_r)(t)=Id(t)$;
	\item Step 5 - The Fuchsian group generators are specified by $S_r, r=1, \cdots, n $, where $n$ is the degree of the hyperelliptic curve;
	\item Step 6 - Fix one of the elliptic transformations, for example, $S_k$, and compute the products $S_kS_r$, for all $r\neq k$;
	\item Step 7 - Check if the transformations $S_kS_r$, calculated in Step 6, are hyperbolic;
	\item Step 8 - The generators of the Fuchsian subgroup are specified by $\{S_kS_r\}$ form the fundamental region.
\end{itemize}

According to \cite{Mursi}, the transformations $S_r(t)$ are expressed by:

$$S_r(t)=\frac{at-e^{\frac{(4n+1)\pi \alpha i}{2}}}{e^{\frac{-(4n+1)\pi \alpha i}{2}}t-a},$$

\noindent where $n=\{0,1,2, \cdots, 2g+1 / 2g+2\}$, $r=\{1,2,3, \cdots, 2g+1 / 2g+2 \}$, $a=(2\cos \pi \alpha -1)^{\frac{-1}{2}}$ and $\alpha = \frac{g-1}{2g+1}$.

\section{Results} \label{Sec3}

In this section, we present the results of this paper, first identifying the minimum and maximum genus associated with the embedding of a $C_{2,8}$ channel and then the characterization process of the Fuchsian group's generators associated with the quantization of this channel, related to genus 2 and 3, detailing and explaining each case performed.

We are interested in a $2$-cell embedding of the complete bipartite graph $K_{2,8}$. Hence, the minimum and the maximum genus of the corresponding surface are given by:

\begin{enumerate}
	\item[$\left( i\right) $]  From (\ref{gen min kmn}), the minimum genus of an oriented compact surface is
	\begin{equation}
		g_{m}\left( K_{2,8}\right) =\{\left( m-2\right) \left(n-2\right)/4\}= \{0\}= 0;
	\end{equation}
	where $\{a\}$ denotes the least integer greater than or equal to the real number $a$.
	
	\item[$\left( ii\right) $]  From (\ref{gen mas kmn}), the maximum genus of an oriented compact surface is
	\begin{equation}
		g_{M}\left( K_{2,8}\right) =\left[ \left( m-1\right) \left(n-1\right) /2\right] = \left[ 3.5 \right]=3.
	\end{equation}
	where $\left[ a\right] $ denotes the greatest integer less than or equal to the real number $a$.
\end{enumerate}

Thus, $g_{m}=0$ and $g_{M}=3$, implying $0\le g\le 3$. We consider the cases where $g=2$ and $g=3$, which are related to the hyperbolic case because $g=0$ and $g=1$, related to the Euclidean case, have already been presented in \cite{canal}.

The identification of the fundamental region through the singularities, in the process of uniformization of the planar algebraic curve associated with the Fuchsian differential equation, will provide, in addition to the surface genus, the generators of the Fuchsian groups associated with the signal constellation, thus enabling the construction and performance analysis of these constellations.

We consider next the two hyperelliptic curves of genus 2, given by $y^2 = z^{5} - 1$, and $y^2 = z^{6} - 1$.
\begin{itemize}
	\item{ \bf{Case 1: $y^2 = z^{5} - 1$}}
	The five singularities may be viewed as the vertices of a regular pentagon, as shown in Figure~\ref{transf1}.
	
	\begin{figure}[!h]
		\centering
		\includegraphics[scale=0.6]{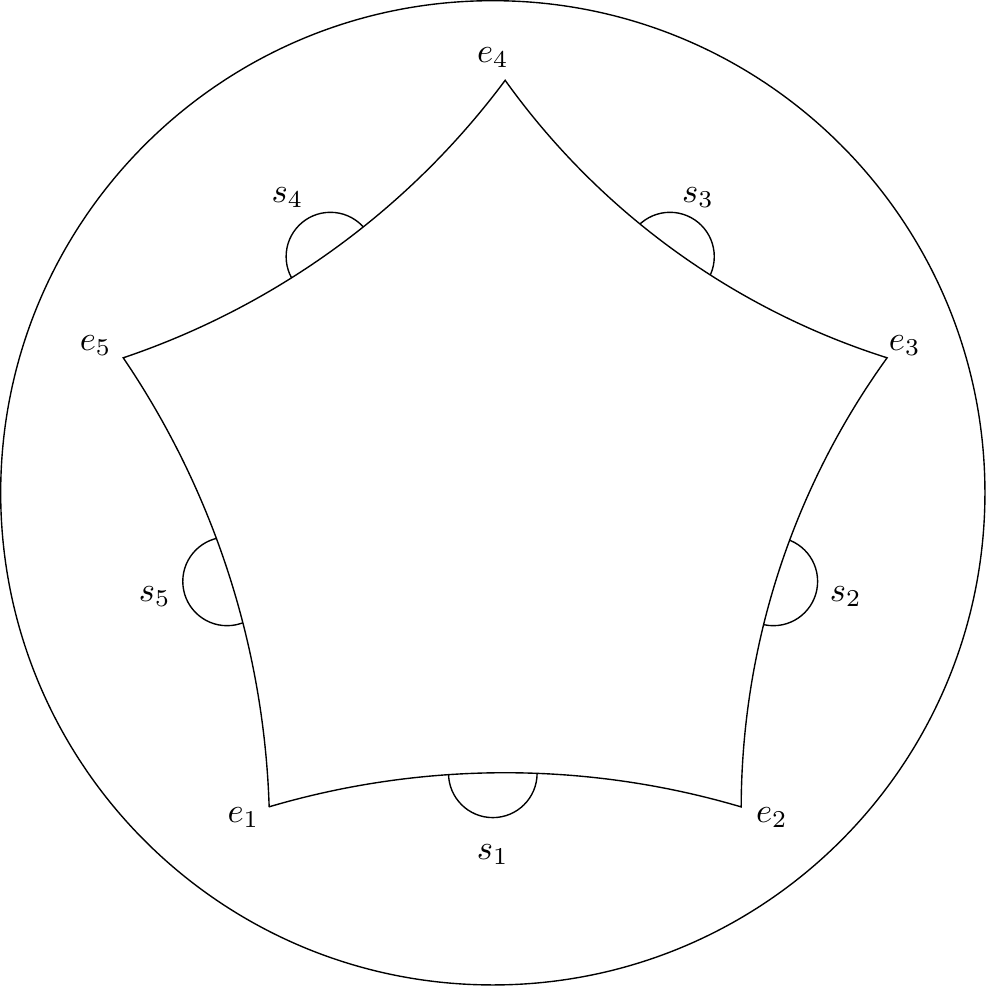}
		\caption{Singularities of $y^2 = z^{5} - 1$.}
		\label{transf1}
	\end{figure}
	
	Note that there is a one-to-one correspondence between the set of solutions of $z^{5}-1$ and the values $-2$, $-1$, $0$, $1$, and $2$, since the roots of the unit divide the circumference into five equal parts, whose angles are $0, 72,144, -72,-144$, and there exists an isomorphism between the multiplicative group  $\exp\{\frac{-i \cdot 2 \cdot 2k\pi}{5}\}$, with $k=0,1,2,3,4$, and the additive group $\mathbb{Z}_5=\{0,1,2,3,4\}$. Hence, the roots (singularities) of the curve $y^2=z^5-1$ may be represented by the elements of the set $\{-2,-1,0,1,2\}$, \cite{canal}. From this fact, the corresponding second-order Fuchsian differential equation is given by:
	
	\begin{small}
		\begin{equation}\label{equ00}
			(z^5-5z^3+4z)y^{\prime\prime}+\left[(z^5-5z^3+4z)\cdot \left(\frac{2}{z+1}+k_1\right)\right]y^{\prime}+[(z^5-5z^3+4z)\cdot k_2]y=0, \quad k_1, k_2 \in \mathbb{C}.
		\end{equation}
	\end{small}
	
	Fixing one of the elliptic transformations shown in  Figure~\ref{transf1}  and multiplying it by the remaining ones results in four-sided pairing hyperbolic transformations (2-tori) as shown in Figure~\ref{transf2}, where $e_i, 1\le i \le 5$, represent the singularities. This polygonal region is the one that uniformizes the given hyperelliptic curve.
	
	\begin{figure}[h!]
		\centering
		\includegraphics[scale=0.6]{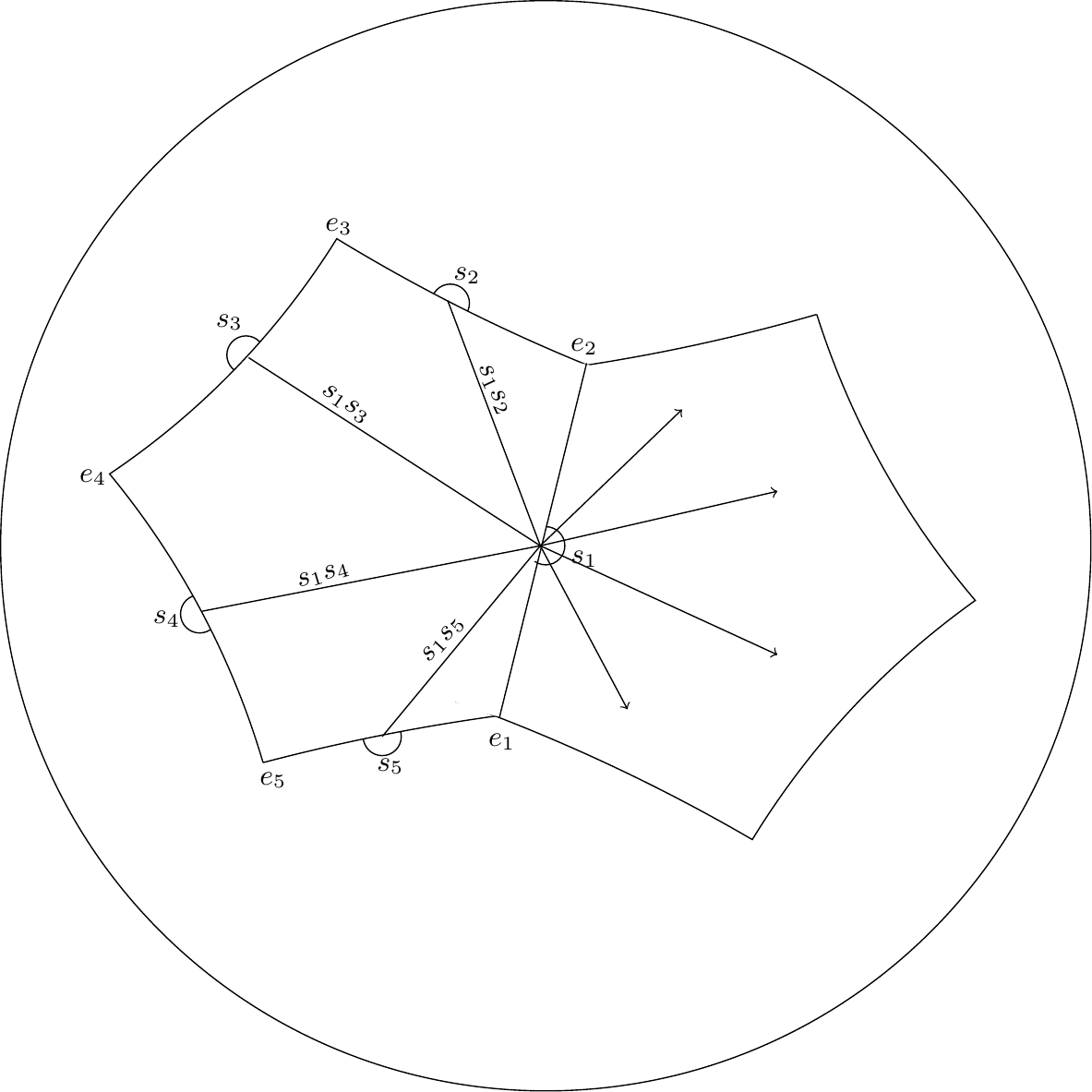}
		\caption{Möbius transformations.}
		\label{transf2}
	\end{figure}
	
	The five singularities are the vertices of the regular hyperbolic polygon. This polygon is the fundamental region of a surface of genus zero whose generators of the Fuchsian group, $\Gamma_0$, are the elliptic transformations. Since the hyperelliptic curve has genus 2, the generators of the corresponding Fuchsian subgroup have to be hyperbolic transformations. In order to obtain such generators, the procedure is to fix one of the elliptic transformations and multiply it by each of the remaining elliptic generators. This procedure leads to the four hyperbolic transformations, as the sequence shows. These four transformations $S_1S_2$, $S_1S_3$, $S_1S_4$, and $S_1S_5$ are the generators of the Fuchsian group $\Gamma_8$ whose fundamental region is the eight-sided regular hyperbolic polygon, 2-tori. Let $T_{i}$, with $i=1,2,3,4$ be such that $T_1 = S_1S_2$, $T_2 = S_1S_3$, $T_3 = S_1S_4$, and $T_4 = S_1S_5$,
	
	\begin{eqnarray*}T_1 = S_1S_2=\left( \begin{array}{ll}
			2.1180371 + 1.538846i & 1.9574413 - 1.4221644i  \\
			1.9574413 + 1.4221644i & 2.1180371 - 1.538846i  \end{array} \right). \end{eqnarray*}
	
	\begin{eqnarray*}T_2 = S_1S_3=\left( \begin{array}{ll}
			3.927059  + 0.9510591i & 3.9148826 - 1.110D-16i  \\
			3.9148826 + 1.110D-16i &  3.927059  - 0.9510591i \end{array} \right). \end{eqnarray*}
	
	\begin{eqnarray*}T_3 = S_1S_4=\left( \begin{array}{ll}
			3.927059  - 0.9510591i & 3.1672066 + 2.3011103i  \\
			3.1672066 - 2.3011103i & 3.927059  + 0.9510591i \end{array} \right). \end{eqnarray*}
	
	\begin{eqnarray*}T_4 = S_1S_5=\left( \begin{array}{ll}
			2.1180371 - 1.538846i & 0.7476761 + 2.3011103i  \\
			0.7476761 - 2.3011103i & 2.1180371 + 1.538846i \end{array} \right). \end{eqnarray*}
	
	The Fuchsian group $\Gamma_8$ has the following presentation:
	$$\Gamma_8 = \langle T_1, T_2, T_3, T_4 : T_1T_2^{-1}T_3T_4^{-1}T_1^{-1}T_2T_3^{-1}T_4 = T_e\rangle ,$$
	where $T_e$ is the identity.
	
	From the hyperelliptic curve, the genus is 2, a 2-tori, for $\partial f=2g+1\Rightarrow 5=2g+1 \Rightarrow g=2$. The matching of this genus with the topological one has to lead the Euler characteristic to a value of 2. Now, the Euler characteristic is given by
	
	\begin{equation}\label{eqq}
		\chi = V-E+F=2-2g.
	\end{equation}
	
	Since the number of vertices is 1, the number of edges is 4, and the number of faces is 1, it follows that $\chi=-2$. Hence, $g=2$. To have $p=8$ sides the fundamental polygon has to be of the form $p=4g$. Thus, the hyperbolic tessellation is $\{4g,q\}$, where $q$ is the number of polygons with $4g$ sides covering each vertex in the tessellation. Since the vertex cycle is $v_c = p/q$ and it equals the number of vertices, that is $v_c = V$, which in this case is $V=1$, it follows that $p=q$. Therefore, the tessellation is $\{4g,4g\}$.
	
	\item {\bf{Case 2: $y^2=z^6-1$}}
	There is a one-to-one correspondence between the six solutions of $z^{6}-1$ and the values $-2$, $-1$, $0$, $1$, $2$, and $3$, since the roots of the unit divide the circumference into six equal parts, as shown in Figure~\ref{4}.

	In order to follow the same procedure as in Case 1, the associated Fuchsian differential equation, the generators of the Fuchsian group, and the tessellation will be presented.
	
	The Fuchsian differential equation is:
	\begin{small}
		\begin{equation*}\label{equ00}
			(z^6-3z^5-5z+4+15z^3+4z^2-12z)y^{\prime\prime}+\left[(z^6-3z^5-5z+4+15z^3+4z^2-12z)\cdot \left(\frac{2}{z-1}+k_1\right)\right]y^{\prime}+
		\end{equation*}
	\end{small}
	
	\begin{small}
		\begin{equation}\label{equ0000}
			+[(z^6-3z^5-5z+4+15z^3+4z^2-12z)\cdot k_2]y=0, \quad k_1, k_2 \in \mathbb{C}.
		\end{equation}
	\end{small}
	
	\begin{figure}[!h]
		\centering
		\includegraphics[scale=0.6]{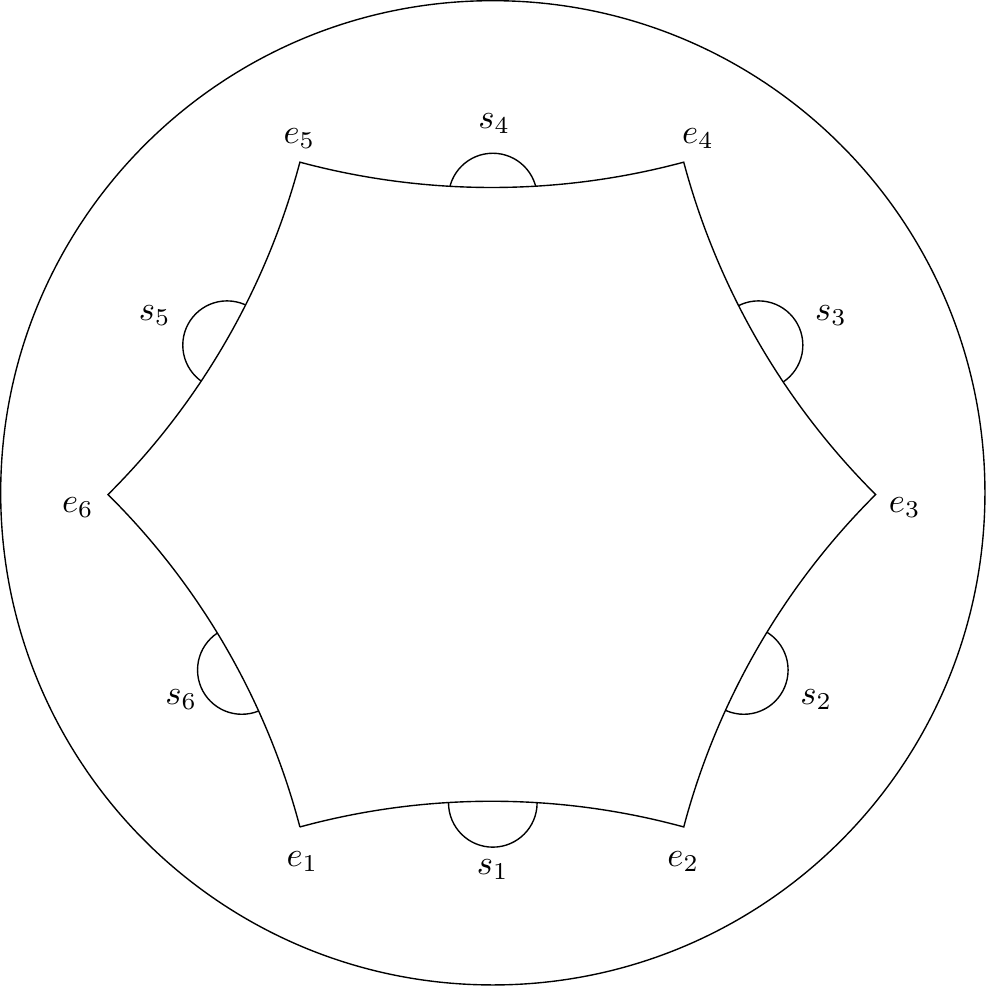}
		\caption{Singularities of $y^2 = z^{6} - 1$.}
		\label{4}
	\end{figure}

	The six singularities are the vertices of the regular hyperbolic polygon. The hyperelliptic curve has genus 2; it follows that the generators of the corresponding Fuchsian subgroup must be hyperbolic transformations. As in Case 1, the procedure is to fix one of the elliptic transformations and multiply it by each of the remaining elliptic generators. This procedure leads to the five hyperbolic transformations as shown in Figure~\ref{5}. These five transformations $S'_1S'_2$, $S'_1S'_3$, $S'_1S'_4$, $S'_1S'_5$, and $S'_1S'_6$ are the generators of the Fuchsian group $\Gamma_{10}$ associated with the fundamental region of the 2-tori. Let $T'_{i}$, with $i=1,2,3,4,5$ be such that $T'_1 = S'_1S'_2$, $T'_2 = S'_1S'_3$, $T'_3 = S'_1S'_4$, $T'_4 = S'_1S'_5$, and $T'_5 = S'_1S'_6$.

	\begin{eqnarray*}T'_1 = S'_1S'_2=\left( \begin{array}{ll}
			2.3660261 + 2.3660267i & 2.2578931 - 2.2578931i  \\
			2.2578931 + 2.2578931i &  2.3660261 - 2.3660267i \end{array} \right). \end{eqnarray*}
	
	\begin{eqnarray*}T'_2 = S'_1S'_3=\left( \begin{array}{ll}
			5.0980784 + 2.3660267i  & 5.3422323 - 1.4314468i  \\
			5.3422323 + 1.4314468i &   5.0980784 - 2.3660267i \end{array} \right). \end{eqnarray*}
	
	\begin{eqnarray*}T'_3 = S'_1S'_4=\left( \begin{array}{ll}
			6.4641046 - 1.110D-16i & 6.1686786 + 1.6528924i  \\
			6.1686786 - 1.6528924i &  6.4641046 + 1.110D-16i \end{array} \right). \end{eqnarray*}
	
	\begin{eqnarray*}T'_4 = S'_1S'_5=\left( \begin{array}{ll}
			5.0980784 - 2.3660267i & 3.9107855 + 3.9107855i  \\
			3.9107855 - 3.9107855i &  5.0980784 + 2.3660267i \end{array} \right). \end{eqnarray*}
	
	\begin{eqnarray*}T'_5 = S'_1S'_6=\left( \begin{array}{ll}
			2.3660261 - 2.3660267i & 0.8264462 + 3.0843393i  \\
			0.8264462 - 3.0843393i &  2.3660261 + 2.3660267i
		\end{array} \right).
	\end{eqnarray*}

\begin{figure}[!h]
	\centering
	\includegraphics[scale=0.8]{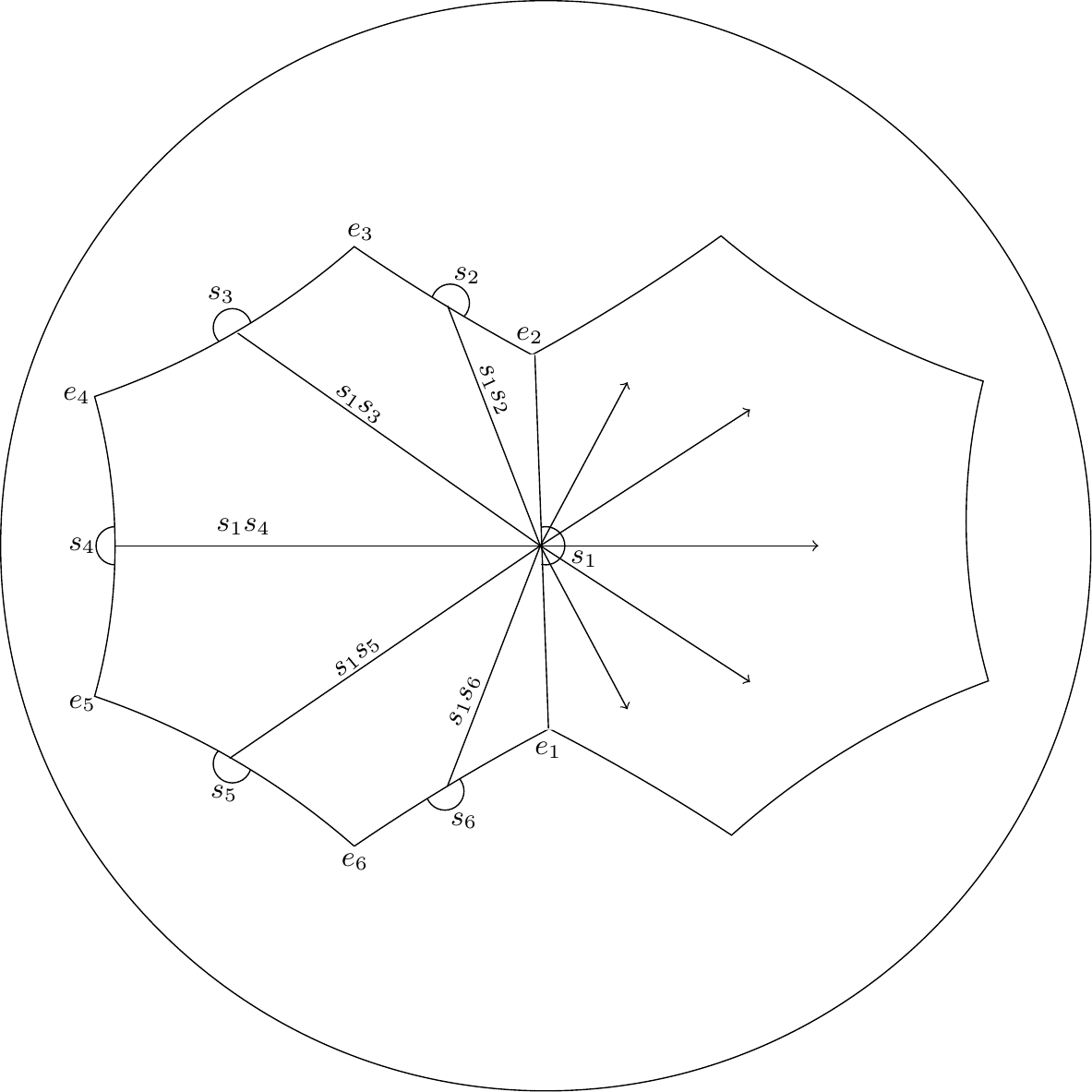}
	\caption{Möbius transformations.}
	\label{5}
\end{figure}
	
	The Fuchsian group $\Gamma_{10}$ has the following presentation:
	$$\Gamma_{10} = \langle T'_1,T'_2,T'_3, T'_4,T'_5 : T'_1{T'_2}^{-1} T'_3 {T'_4}^{-1} {T'_5}= {T'_1}^{-1} {T'_2} {T'_3}^{-1} {T'_4} {T'_5}^{-1}=T_e\rangle ,$$
	where $T_e$ is the identity.
	
	From the hyperelliptic curve, the genus is 2, for $\partial f=2g+2\Rightarrow 6=2g+2 \Rightarrow g=2$. As in Case 1, to match the value of this genus with the topological one, the Euler characteristic has to lead to a value of 2. In this case, the fundamental region has ten sides. Now, the Euler characteristic, as given by (\ref{eqq}), leads to $\chi = V-E+F = 2-5+1=-2$, where the number of vertices is 2, the number of edges is 5, and the number of faces is 1, it follows that $\chi=-2$. Hence, $g=2$. To have $p=10$ sides, the fundamental polygon has to be given by $p=4g+2$. Thus, the hyperbolic tessellation is $\{4g+2,q\}$, where $q$ is the number of polygons with $4g+2$ sides covering each vertex in the tessellation. Since the vertex cycle is $v_c = p/q$ and it equals the number of vertices, that is, $v_c = V$, which in this case is $V=2$, it follows that $p=2q$. Therefore, the tessellation is $\{4g+2,2g+1\}$.
\end{itemize}

From the analysis of Cases 1 and 2 it follows that:
\begin{itemize}
	\item[1)] the same genus $g=2$ is obtained using two distinct hyperelliptic curves $y^2=z^5-1$ and $y^2=z^6-1 $;
	
	\item[2)] $\Gamma_8$ and $\Gamma_{10}$ have different generators implying that they are not isomorphic;
	
	\item[3)]$\Gamma_8$ and $\Gamma_{10}$ lead to different tessellations, $\{8,8\}$ for Case 1, and $\{10,5\}$ for Case 2, respectively;
	
	\item[4)] the differences found from the computations associated with the different tessellations lead to different possibilities of applications and performance analysis in the construction of codes and applications involving the different tessellations related to the same genus (one being denser than the other, for instance);
	
	\item[5)] Lastly, the hyperbolic areas associated with the uniformization regions of the curves $z^5-1$ and $z^6-1$ are the same, whereas the number of neighbors is not. Consequently, a tessellation with more neighbors leads to a slightly higher error probability.
\end{itemize}

Finally, consider the hyperelliptic curves with genus 3, given by $y^2 = z^{7} - 1$, and $y^2 = z^{8} - 1$.
\begin{itemize}
	\item {\bf{Case 3: $y^2 = z^{7} - 1$ }}
	Note that these seven singularities may be viewed as the vertices of a regular heptagon, as shown in Figure~\ref{6}.
	
	\begin{figure}[!h]
		\centering
		\includegraphics[scale=0.7]{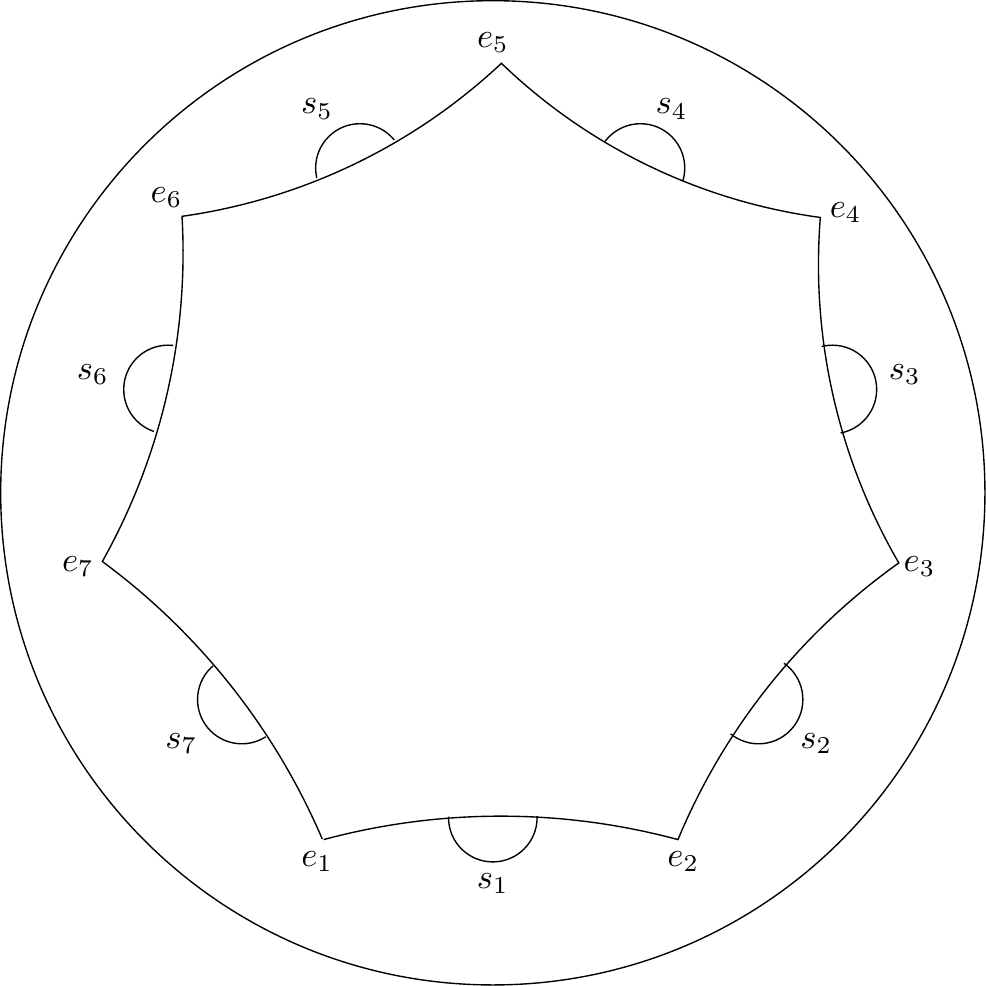}
		\caption{Singularities of $y^2 = z^{7} - 1$.}
		\label{6}
	\end{figure}
	
	There is a one-to-one correspondence between the set of solutions of $z^{7}-1$ and the values $-3$, $-2$, $-1$, $0$, $1$, $2$ and $3$, since the roots of the unit divide the circumference into seven equal parts. In order to follow the same procedure as in Case 1 and Case 2, the associated Fuchsian differential equation, the generators of the Fuchsian group, and the tessellation is considered.
	
	The Fuchsian differential equation is given by:
	
	\begin{small}
		\begin{equation*}\label{equ00}
			(z^7-14z^5+49z^3-36z)y^{\prime\prime}+\left[(z^7-14z^5+49z^3-36z)\left(\frac{2}{z+1}+k_1\right)\right]y^{\prime}+
		\end{equation*}
	\end{small}
	
	\begin{small}
		\begin{equation}\label{equ00}
			+ \left[(z^7-14z^5+49z^3-36z).k_2]y\right]=0, \quad k_1, k_2 \in \mathbb{C}.
		\end{equation}
	\end{small}
	
	The seven singularities are the vertices of the regular hyperbolic polygon. Since the hyperelliptic curve has genus 3, the generators of the corresponding Fuchsian subgroup have to be hyperbolic transformations. In order to obtain such generators, the procedure is to fix one of the elliptic transformations and multiply it by each of the remaining elliptic generators, as shown in Figure~\ref{7}. This procedure leads to the six hyperbolic transformations, as shown next. These six transformations $S''_1S''_2$, $S''_1S''_3$, $S''_1S''_4$, $S''_1S''_5$, $S''_1S''_6$, $S''_1S''_7$ are the generators of the Fuchsian group $\Gamma_{12}$ associated with the fundamental region of the 3-tori. Let $T''_{i}$, with $i=1,2,3,4,5$ be such that $T''_1 = S''_1S''_2$, $T''_2 = S''_1S''_3$, $T''_3 = S''_1S''_4$, $T''_4 = S''_1S''_5$, $T''_5 = S''_1S''_6$, and $T''_6 = S''_1S''_7$.
	
	\begin{figure}[!h]
		\centering
		\includegraphics[scale=1.0]{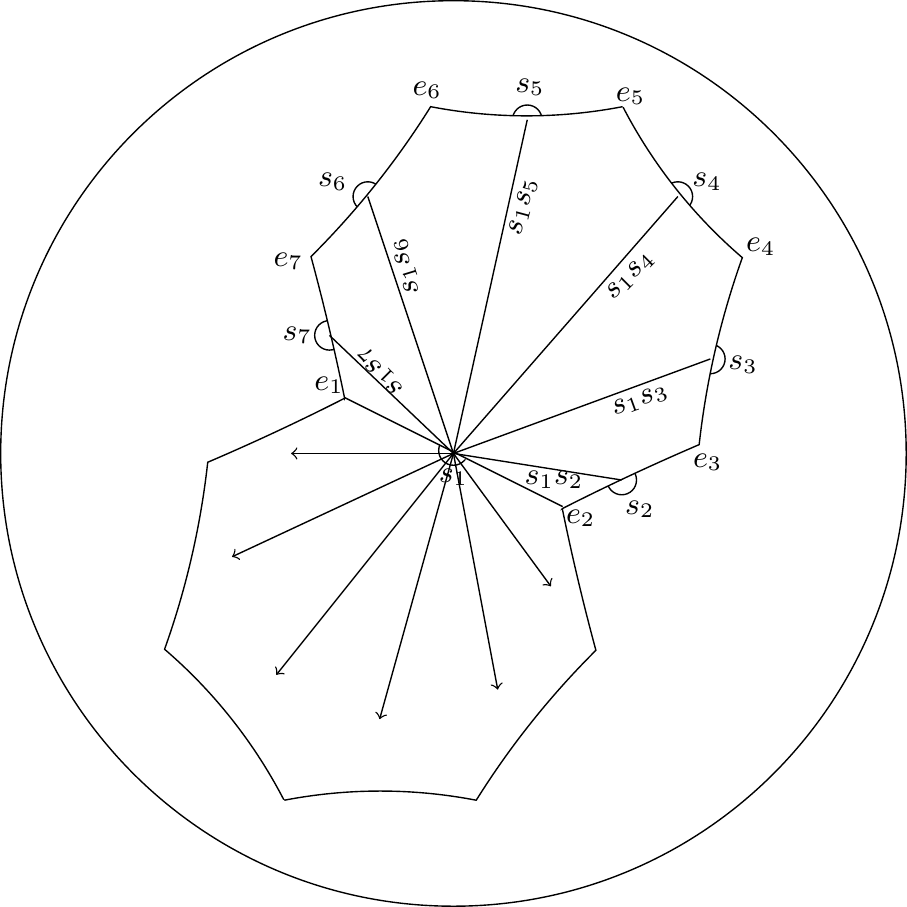}
		\caption{Möbius transformations.}
		\label{7}
	\end{figure}
	
	\begin{eqnarray*}T''_1 = S''_1S''_2=\left( \begin{array}{ll}
			1.400969  + 0.3197621i & 1.0060962 - 0.2296349i  \\
			1.0060962 + 0.2296349i &  1.400969  - 0.3197621i \end{array} \right). \end{eqnarray*}
	
	\begin{eqnarray*}T''_2 = S''_1S''_3=\left( \begin{array}{ll}
			1.62349   - 0.1423075i  &  1.0060962 + 0.802335i  \\
			1.0060962 - 0.802335i &  1.62349   + 0.1423075i \end{array} \right). \end{eqnarray*}
	
	\begin{eqnarray*}T''_3 = S''_1S''_4=\left( \begin{array}{ll}
			1.1234898 - 0.2564293i &  -2.220D-16 + 0.5727001i  \\
			-2.220D-16 - 0.5727001i &  1.1234898 + 0.2564293i \end{array} \right). \end{eqnarray*}
	
	\begin{eqnarray*}T''_4 = S''_1S''_5=\left( \begin{array}{ll}
			1.1234898 + 0.2564293i & 0.447755  - 0.3570727i  \\
			0.447755  + 0.3570727i &  1.1234898 - 0.2564293i \end{array} \right). \end{eqnarray*}
	
	\begin{eqnarray*}T''_5 = S''_1S''_6=\left( \begin{array}{ll}
			1.62349   + 0.1423075i &  1.2545815 + 0.28635i  \\
			1.2545815 - 0.28635i &  1.62349   - 0.1423075i
		\end{array} \right). \end{eqnarray*}
	
	\begin{eqnarray*}T''_6 = S''_1S''_7=\left( \begin{array}{ll}
			1.400969 - 0.3197621i &  0.447755 + 0.9297727i  \\
			0.447755 - 0.9297727i &   1.400969 + 0.3197621i \end{array} \right).
	\end{eqnarray*}
	
	The Fuchsian group $\Gamma_{12}$ has the following presentation: $$\Gamma_{12} = \langle T''_1, T''_2, T''_3, T''_4, T''_5T''_5,T''_6 : T''_1{T''_2}^{-1} T''_3 {T''_4}^{-1} {T''_5}{T''_6}^{-1}{T''_1}^{-1} {T''_2} {T''_3}^{-1} {T''_4} {T''_5}^{-1}{T''_6}=T_e\rangle ,$$
	where $T_e$ is the identity.
	
	From the given hyperelliptic curve, the genus is 3, a 3-tori, for $\partial f=2g+1\Rightarrow 7=2g+1 \Rightarrow g=3$. As in the previous cases, to match the value of this genus with the topological one, the Euler characteristic has to lead to a value of 3. Note that in this case, the fundamental region has 12 sides. Now, the Euler characteristic, as given by (\ref{eqq}), leads to $\chi = V-E+F = 1-6+1=-4$, where the number of vertices is 1, the number of edges is 6, and the number of faces is 1, so $\chi=-4$. Hence, $g=3$. To have $p=12$ sides, the fundamental polygon has to be given by $p=4g$. Thus, the hyperbolic tessellation is $\{4g,q\}$, where $q$ is the number of polygons with $4g$ sides covering each vertex in the tessellation. Since the vertex cycle is $v_c = p/q$ and it equals the number of vertices, that is, $v_c = V$, which in this case is $V=1$, it follows that $p=q$. Therefore, the tessellation is $\{4g,4g\}$. Therefore, the tessellation is $\{12,12\}$.
	
	\item {\bf {Case 4: $y^2=z^8-1$ }}
	There exists a bijective correspondence among the eight solutions of $z^{8}-1$ and the values $-3$, $-2$, $-1$, $0$, $1$, $2$, $3$ and $4$, since the roots of the unit divide the circumference into eight equal parts, as shown in Figure~\ref{8}.
	
	The Fuchsian differential equation is given by:
	
	\begin{small}
		\begin{equation*}\label{equ00}
			(z^8-4z^7-14z^6+56z^5+49z^4-196z^3-36z^2+144z)y^{\prime\prime}+\left[(z^8-4z^7-14z^6+56z^5+49z^4-196z^3-36z^2+144z)\right].
		\end{equation*}
	\end{small}
	
	\begin{small}
		\begin{equation}\label{equ00000}
			\cdot \left(\frac{2}{z-1}+k_1\right)y^{\prime}+\left[(z^8-4z^7-14z^6+56z^5+49z^4-196z^3-36z^2+144z)\cdot k_2\right]y=0, \quad k_1, k_2 \in \mathbb{C}.
		\end{equation}
	\end{small}
	
	\begin{figure}[!h]
		\centering
		\includegraphics[scale=0.8]{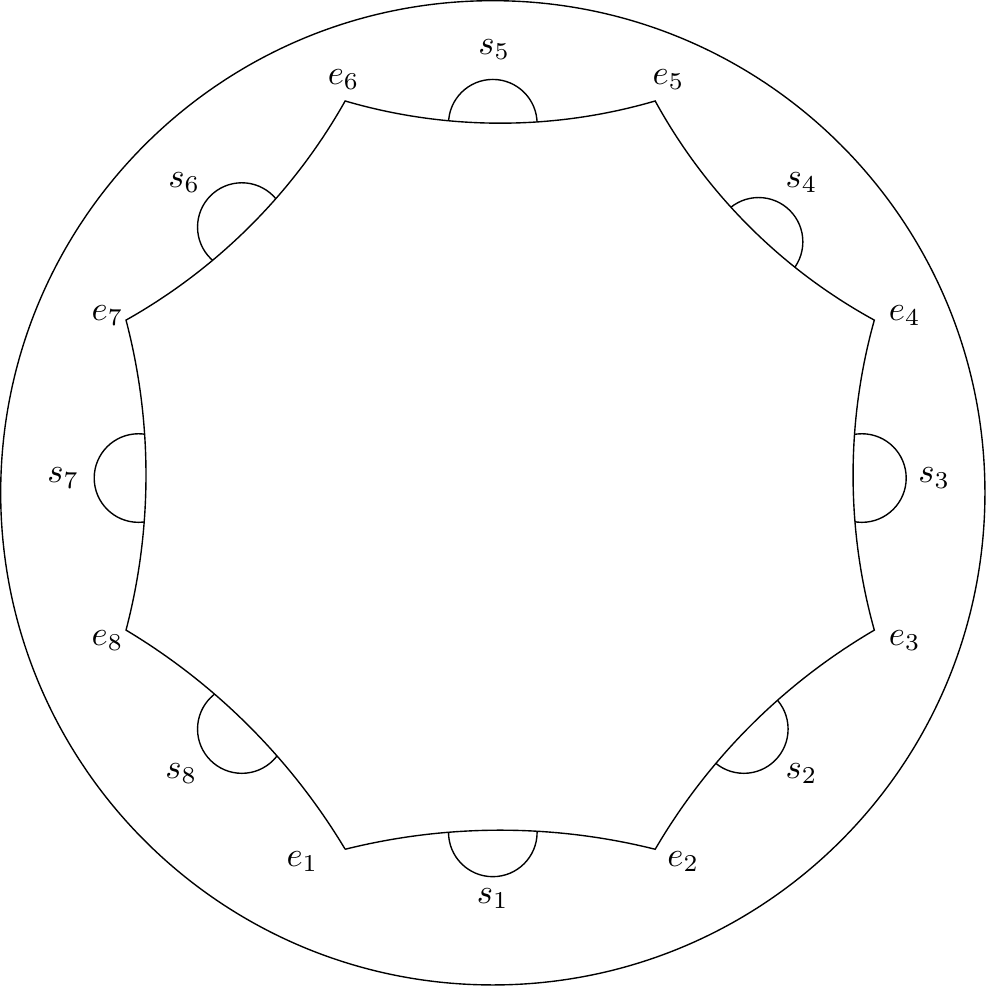}
		\caption{Singularities of $y^2 = z^{8} - 1$.}
		\label{8}
	\end{figure}
	
	The eight singularities are the vertices of the regular hyperbolic polygon. Since the hyperelliptic curve has genus 3, the generators of the corresponding Fuchsian subgroup have to be hyperbolic transformations. We need to fix one of the elliptic transformations and multiply it by each of the remaining elliptic generators, as shown in Figure~\ref{9}. This procedure leads to the seven hyperbolic transformations, as shown next. These seven transformations $S^*_1S^*_2$, $S^*_1S^*_3$, $S^*_1S^*_4$, $S^*_1S^*_5$, $S^*_1S^*_6$, $S^*_1S^*_7$, $S^*_1S^*_8$ are the generators of the Fuchsian group $\Gamma_{14}$ associated with the fundamental region of the 3-tori. Let $T^*_{i}$, with $i=1,2,3,4,5,6,7$ be such that $T^*_1 = S^*_1S^*_2$, $T^*_2 = S^*_1S^*_3$, $T^*_3 = S^*_1S^*_4$, $T^*_4 = S^*_1S^*_5$, $T^*_5 = S^*_1S^*_6$, $T^*_6 = S^*_1S^*_7$, and $T^*_7 = S^*_1S^*_8$.
	
	\begin{eqnarray*}T^*_1 = S^*_1S^*_2=\left( \begin{array}{ll}
			1.7071069 + 0.7071069i & 1.4355002 - 0.5946036i  \\
			1.4355002 + 0.5946036i  &  1.7071069 - 0.7071069i   \end{array} \right). \end{eqnarray*}
	
	\begin{eqnarray*}T^*_2 = S^*_1S^*_3=\left( \begin{array}{ll}
			2.4142138 + 5.551D-17i  &  2.0301038 + 0.8408965i\\
			2.0301038 - 0.8408965i  &   2.4142138 - 5.551D-17i \end{array} \right). \end{eqnarray*}
	
	\begin{eqnarray*}T^*_3 = S^*_1S^*_4=\left( \begin{array}{ll}
			1.7071069 - 0.7071069i &  0.5946036 + 1.4355002i \\
			0.5946036 - 1.4355002i  &  1.7071069 + 0.7071069i  \end{array} \right). \end{eqnarray*}
	
	\begin{eqnarray*}T^*_4 = S^*_1S^*_5=\left( \begin{array}{ll}
			1. + 1.110D-16i & 0. - 2.220D-16i \\
			0. + 2.220D-16i &  1. - 1.110D-16i \end{array} \right). \end{eqnarray*}
	
	\begin{eqnarray*}T^*_5 = S^*_1S^*_6=\left( \begin{array}{ll}
			1.7071069 + 0.7071069i  & 1.4355002 - 0.5946036i  \\
			1.4355002 + 0.5946036i & 1.7071069 - 0.7071069i \end{array} \right). \end{eqnarray*}
	
	\begin{eqnarray*}T^*_6 = S^*_1S^*_7=\left( \begin{array}{ll}
			2.4142138 + 0.i  &  2.0301038 + 0.8408965i  \\
			2.0301038 - 0.8408965i &  2.4142138 + 0.i \end{array} \right). \end{eqnarray*}
	
	\begin{eqnarray*}T^*_7 = S^*_1S^*_8=\left( \begin{array}{ll}
			1.7071069 - 0.7071069i &  0.5946036 + 1.4355002i  \\
			0.5946036 - 1.4355002i & 1.7071069 + 0.7071069i  \end{array} \right). 
	\end{eqnarray*}
	
	The Fuchsian group $\Gamma_{14}$ has the following presentation:
	$$\Gamma_{14} = \langle T^*_1, T^*_2, T^*_3, T^*_4, T^*_5, T^*_6, T^*_7  : T^*_1{T^*_2}^{-1} T^*_3 {T^*_4}^{-1} {T^*_5}{T^*_6}^{-1}{T^*_7} = {T^*_1}^{-1} {T^*_2} {T^*_3}^{-1} {T^*_4} {T^*_5}^{-1}{T^*_6}{T^*_7}^{-1} = T_e\rangle ,$$
	where $T_e$ is the identity.
	
	\begin{figure}[!h]
	\centering
	\includegraphics[scale=1.0]{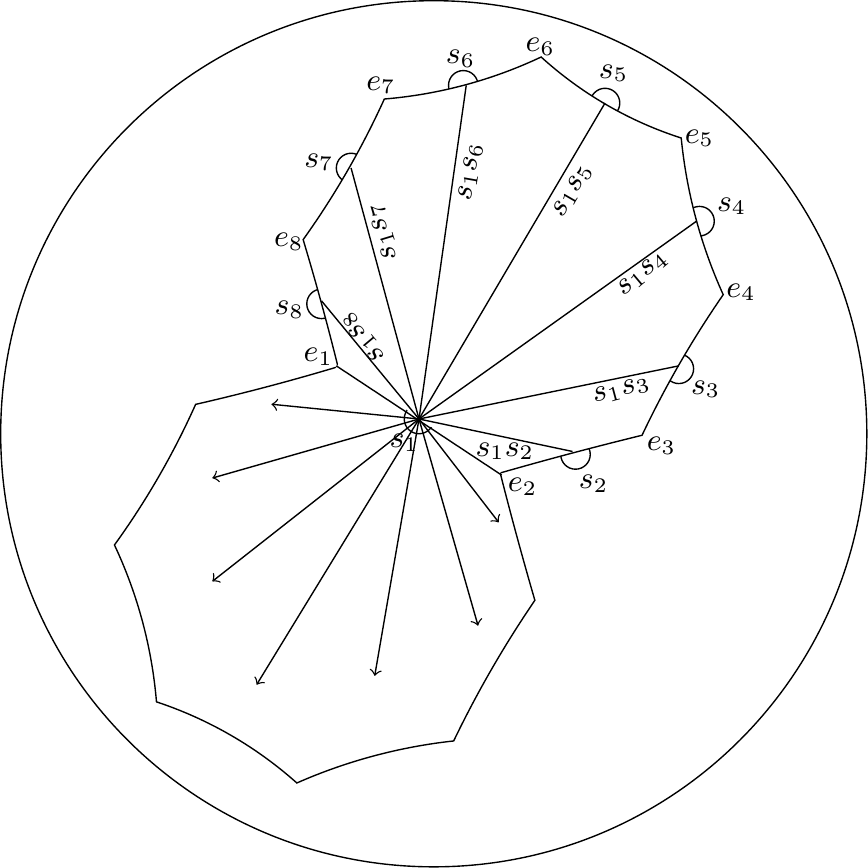}
	\caption{Möbius transformations.}
	\label{9}
	\end{figure}
	
	From the given hyperelliptic curve, the genus is 3, a 3-tori, for $\partial f=2g+2\Rightarrow 8=2g+2 \Rightarrow g=3$. As in the previous cases, to match the value of this genus with the topological one, the Euler characteristic has to lead to a value of 3. Note that in this case, the fundamental region has 14 sides. From the Euler characteristic, given by (\ref{eqq}), $\chi = V-E+F = 2-7+1=-4$, where the number of vertices is 2, the number of edges is 7, and the number of faces is 1, it follows that $\chi=-4$. Hence, $g=3$. To have $p=14$ sides, the fundamental polygon has to be given by $p=4g+2$. Thus, the hyperbolic tessellation is $\{4g+2,q\}$, where $q$ is the number of polygons with $4g+2$ sides covering each vertex in the tessellation. Since the vertex cycle is $v_c = p/q$ and it equals the number of vertices, that is, $v_c = V$, which in this case is $V=2$, it follows that $p=2q$. Therefore, the tessellation is of the form $\{4g+2,2g+1\}$, leading to the $\{14,7\}$ tessellation.
\end{itemize}

From the analysis of Cases 3 and 4 it follows that:
\begin{itemize}
	\item[1)] The same genus $g=3$ is obtained using two distinct hyperelliptic curves $y^2=z^7-1$ and $y ^2=z^8-1$;
	
	\item[2)] $\Gamma_{12}$ and $\Gamma_{14}$ have different generators implying that they are not isomorphic;
	
	\item[3)] $\Gamma_{12}$ and $\Gamma_{14}$ lead to different tessellations, $\{12,12\}$ for Case 3, and $\{14,7\}$ for Case 4;
	
	\item[4)] the differences found from the computations associated with the different tessellations lead to different possibilities of applications and performance analysis in the construction of codes and applications involving the different tessellations related to the same genus (one being denser than the other, for instance);
	
	\item[5)] Lastly, the hyperbolic areas associated with the uniformization regions of the hyperelliptic curves $z^7-1$ and $z^8-1$ are the same, but the number of neighbors is different. Consequently, a tessellation with more neighbors leads to a slightly higher error probability.
\end{itemize}

\section{Final Remarks} \label{Sec4}
In this paper, we have considered the steps to be followed to analyze the algebraic and geometric characterizations related to the channel
quantization $C_{2,8}$. Initially, embedding the complete bipartite graph $K_{2,8}$ associated with the channel $C_{2,8}$ in compact surfaces was analyzed, whose genus is in the interval $0\leq g \leq 3 $. The hyperbolic cases ($g=2$ and $g=3$) were considered because the cases for $g=0$ and $g=1$ have already been presented in \cite{canal}. The hyperelliptic curves were analyzed and subdivided into two cases: 1) $g=2$, for the curves $y^2=z^5-1$ and $y^2=z^6-1$; and 2) $g=3$, for the curves $y^2=z^7-1$ and $y^2=z^8-1$. Next, the Fuchsian differential equations associated with each of the cases were identified in order to analyze, through the singularities of these equations, the fundamental polygon associated with the process and the possible transformations of side pairings, as proposed by \cite{Whittaker, Whittaker2, Mursi}, in order to establish the uniformization region of the corresponding hyperelliptic curve. Additionally, the tessellations associated with each of the cases were identified: for $g=2$, the tessellations $\{8,8\}$ and $\{10,5\}$ and for $g=3$, the tessellations $\{12,12\}$ and $\{14,7\}$. Note that for the same genus, in addition to different Fuchsian group generators (non-isomorphic), different tessellations were obtained. These differences are relevant because they lead to different applications and performance analysis possibilities in constructing codes and applications involving different tessellations related to the same genus and with different fundamental regions. Such a difference will be significant in different situations involving the design of digital communication systems. The results can help engineers, cryptographers, mathematicians, and physicists, among others, use these mathematical tools in constructing and analyzing new, more reliable, and less complex communication systems.

\section*{Acknowledgement}
This work is supported by CNPq under grant 305239/2020-1 and FAPEMIG under grant APQ-00019-21.

\small{

}

\end{document}